\documentclass[graybox]{svmult}

\usepackage{cite}
\usepackage{mathptmx}       
\usepackage{helvet}         
\usepackage{courier}        
\usepackage{type1cm}        
\usepackage{makeidx}         
\usepackage{graphicx}        
\usepackage{multicol}        
\usepackage[bottom]{footmisc}

\makeindex             

\begin{document}

\title*{Classical and quantum: a conflict of interest}
\author{T. P. Singh}
\institute{T. P. Singh \at Tata Institute of Fundamental Research, Homi Bhabha Road, Mumbai 400005, India\\\ email: {tpsingh@tifr.res.in}}
%
%
\maketitle


\abstract{We highlight three conflicts between quantum theory and classical general relativity, which make it implausible that  a  quantum theory of gravity can be arrived at by quantising classical gravity. These conflicts are: quantum nonlocality and space-time structure; the problem of time in quantum theory; and the quantum measurement problem. We explain how these three aspects bear on each other, and how they point towards an underlying noncommutative geometry of space-time.  }

\bigskip

{\it \noindent This article is warmly dedicated to my Ph. D. supervisor Thanu Padmanabhan, on the happy occasion of his sixtieth birthday.
One of the most important things I learnt from Paddy was to look for one's own questions and one's own answers, instead of necessarily accepting someone else's  line of thought. I hope this lesson is reflected in the ideas presented in this article. In particular, Paddy himself might not agree with some or many of these ideas, and in that sense the lesson has probably been learnt well!}

\section{Some limitations of quantum theory}
\label{sec:1}
Quantum theory is extraordinarily successful, and is not contradicted by any experiment. This is true for its non-relativistic version, as well as  for relativistic quantum mechanics, and for quantum field theory. However, its successes should not blind us to the limitations of its theoretical structure, as we understand it today.  First and foremost though, it is important to remember, and not often emphasized, that quantum mechanics has not been tested in all parts of the parameter space that are in principle accessible in table-top laboratory experiments. We have in mind tests of  quantum linear superposition [Schrodinger cat states] for mesoscopic objects. The largest objects for which the superposition principle has been tested have a mass of about $10^5$ a.m.u. and the smallest objects which are known to behave classically have a mass of about a microgram [i.e. about $10^{18}$ a.m.u.]. In between, there is a technologically challenging range of some thirteen orders of magnitude, where there are no experimental tests of the superposition principle, although significant progress is now taking place since the last few years. [We note that macroscopic superpositions of internal states as in superconductors and Bose-Einstein condensates do not negate the previous statement. More on this later.] In this untested intermediate range, maybe there is a quantum-to-classical transition which can be explained by environmental decoherence and the many-worlds interpretation, or maybe by Bohmian mechanics. Alternatively, it maybe the case that there is a new dynamics such as spontaneous collapse, to which quantum and classical mechanics are approximations, and whose effects become significant in this intermediate regime, and which is responsible for the quantum-to-classical transition. To believe that quantum mechanics will definitely not be violated in this yet untested regime is akin to believing, if one were in the nineteenth century, that Newton mechanics will not be violated at high speeds or for small objects, even though the theory was not then tested at high speeds or for small objects. Of course with hindsight we know that such faith in Newtonian mechanics was misplaced, and accordingly we should reserve our judgement about quantum mechanics as well, until these thirteen orders of magnitude have been covered by experiments. 

Quantum mechanics is generally taught to students as a `final' theory, with rarely a mention of the unsatisfying aspects of its theoretical construction. Many physicists painfully `unlearn' the theory in their later years, and realise the extreme peculiarity of the structure of the theory. The strangest aspect is the extreme dependence of the theory on its own classical limit, for its very construction and interpretation. One starts from the classical [Lagrangian or Hamiltonian] dynamics of the theory for the chosen degrees of freedom, and one must know the classical action and the Poisson brackets. Then the peculiar procedure `quantize' is invoked: configuration variables and their canonical momenta are raised to the level of operators, and Poisson brackets are replaced by ad hoc quantum commutation relations. It works perfectly, but one is left wondering if the construction is fundamental: one should have been able to write  down the principles of quantum theory ab initio, and derive classical mechanics from them, rather than the other way round.

The dependence on classical limit continues when one faces the task of interpreting the results of experiments on quantum systems, giving rise to the infamous quantum measurement problem ~\cite{Wheeler-Zurek:1983}. There is a need for a so-called classical measuring apparatus: an object which is not found in superposition of position states, so that classical pointer states [which define the outcome of a measurement] can be defined. But then we are faced with tough questions. How large should an object be before it can be called classical? Quantum mechanics is silent about this. And the classical apparatus  which quantum mechanics so much depends on for its interpretation, is something whose classical properties [in particular, the absence of position superposition of pointer states] should have been derived  from quantum mechanics, rather than assuming its existence a priori, as if it had nothing to do with quantum theory per se.

It is well-known of course that things get more difficult from this point on. The evolution of the state of the quantum system is described by the Schr\"{o}dinger equation: this evolution is deterministic and linear. The process of measurement by the classical apparatus breaks both linear superposition and determinism. Although there is no randomness in the initial conditions for the Schr\"{o}dinger evolution, the outcomes of the measurement are random and probabilistic. This is an unparalleled situation in physics: probabilities without random initial conditions. The fact that probabilities arise during measurement, implies that something has to give. It means that either the probabilities are not real but only apparent, or that there is an aspect of randomness in the dynamics, or in the initial conditions, which is not evident in the Schr\"{o}dinger equation. 

Not only is there a dependence on its own classical limit, but there is also a dependence of quantum theory on external spacetime structure. We emphasize two aspects of this: one which suggests a possible conflict with special relativity, and the other which strongly suggests that the present formulation of the theory should possess an equivalent, but a more fundamental, formulation. The first of these has to do with the EPR paradox and non-local quantum correlations, which suggest that quantum events influence each other outside the light cone. One possible implication of this is that wave-function collapse in quantum theory 
is simply not compatible with the spacetime structure dictated by special relativity, and  in order to describe collapse satisfactorily one perhaps  needs to introduce a new `quantum' structure of spacetime. 

The second aspect, rarely emphasized, has to do with the fact that the time that appears in quantum theory is part of a classical spacetime geometry, which geometry is produced by classical macroscopic objects. But these classical objects are in turn a limiting case of quantum theory!  Once again, the dependence of the theory on its own limit is evident. Clearly, there then ought to exist an equivalent reformulation of quantum theory which does not refer to a classical time.

We thus see that there are at least three different ways in which quantum mechanics depends on its own classical limit, or on classical spacetime structure.  These give rise to the quantum measurement problem, the problem of quantum nonlocality, and the problem of time in quantum theory. In the next three sections we briefly review some developments which address these problems, and their inter-relationship. In the last section we discuss what these problems and their possible resolutions imply for a future quantum theory of gravity.

\section{The quantum measurement problem}
\label{sec:2}
Modern approaches to addressing the measurement problem broadly fall into three classes. The first is to say that collapse of the wave function is only an apparent process, and in reality no collapse ever takes place This is the essence of the many worlds interpretation. There is no need to modify or reformulate quantum theory. The second is to say that there is randomness in the initial conditions, but the evolution by itself is deterministic. This is Bohmian mechanics - a mathermatical reformulation of quantum mechanics. The third is to say that there is randomness in the dynamics, and the deterministic  Schr\"{o}dinger evolution is only an approximation to the random dynamics. This is the essence of collapse models. 

According to the many worlds interpretation, the evolution is deterministic Schr\"{o}dinger evolution through and through, and upon a measurement the universe, including the observer, `splits' into many branches, with a given branch possessing only one out of the various possible outcomes. The other branches contain, respectively, one or the other outcomes. The different branches do not interfere with each other, presumably because of decoherence. [There is a vast literature on decoherence, including the experiments and models by ~\cite{Brune:96, Harris:81,Gerlich2007kapitza},   books by  ~\cite{Joos:03, Schlosshauer:2007, Breuer:2000} and the seminal
papers ~\cite{Zeh:70, Joos:85, Caldeira:81}  and ~reviews
~\cite{Schlosshauer:2005, Vacchini:2009, Zurek:91, Zurek:03}, and ~\cite{Bacciagaluppi:2007}.] The collapse of the wave function is only apparent, not real, and there is no need to modify quantum mechanics. The hard part about many worlds is to understand where the probabilities come from? If the evolution is always deterministic Schr\"{o}dinger evolution, then why do the outcomes obey the Born probability rule?  Various explanations have been put forward, but they do not appear convincing enough ~\cite{Everett:57, DeWitt:73, Tegmark:2007, Kent:1990, Wallace:2003, Deutsch:1998, Vaidman:2002,Hsu:2011, Saunders2010, Putnam:05, Barrett:12, Bacciagaluppi:2001}.

Bohmian mechanics is a neat and precise reformulation of quantum theory, where additional equations of motion are introduced for the positions of particles. The wave function, which satisfies the Schr\"{o}dinger equation, also enters in the equation of motion of particles. The theory is a deterministic theory of particles in motion. Randomness enters in a classical sense, via random initial conditions, chosen such that the outcomes of experiments obey the Born rule. Bohmian mechanics, as well as many worlds, make the same predictions as quantum theory, and they would be falsified if collapse models, which predict departures from quantum theory, are experimentally verified ~\cite{Bohm:52, Bohm2:52, Bub:1997, Duerr:92, Holland, Bohmbook, Duerr, DGZ}.

Collapse models, first developed in the eighties,  propose a stochastic, nonlinear modification of the Schr\"{o}dinger equation, and introduce the new feature that collapse of the wave function is a spontaneous process, not having anything to do per se with the act of measurement ~\cite{Pearle:76, Ghirardi:86, Ghirardi2:90, Pearle:99, Bassi:03}. There is no longer any need for the vaguely defined measuring apparatus, nor an artificial divide between a `quantum system' and a `classical apparatus'.  The nonlinear modification breaks linear superposition, while its stochastic nature ensures that the outcome of the broken superposition is random. The structure of the modifying terms is chosen in such a way that the random outcomes are realised according to the Born probability rule. The theory introduces two new constants of nature, a rate constant $\lambda$ which determines the rate of collapse, and a critical length $r_c$ to which the collapsed wave function is confined. The rate constant has been assigned an ad hoc value of $10^{-17}$ sec$^{-1}$ 
for a nucleon - this means that the wave function of a nucleon undergoes spontaneous collapse once every $10^{17}$ sec. Understandably then, the nonlinear modification is completely negligible for the nucleon and it behaves perfectly quantum mechanically, obeying the Schr\"{o}dinger equation. However, for a particle of mass $m$, the rate constant is assumed to be $(m/m_{N})\lambda$, where $m_N$ is the nucleon mass, and hence the rate constant scales with mass. For macroscopic objects, the wave function collapses extremely rapidly; this explains the classical nature of macroscopic objects, and in particular it explains why pointer position states are classical. 

Collapse models thus also provide a natural solution to the measurement problem. Before a quantum system interacts with the measuring apparatus, its microscopic nature ensures that the rate constant is very small, and the superpositions are long lived. Upon its interaction with the so-called measuring apparatus [which is macroscopic] their entangled state represents a macroscopic superposition, which involves the superposition of pointer position states. This state is extremely short lived, according to the model, and very quickly `collapses' to one of the outcomes, while obeying the Born rule. 

These models propose that there is a new stochastic dynamics, to which quantum mechanics is the microscopic approximation, and classical mechanics is the macroscopic approximation. The stochastic effect is negligible in the microscopic limit. On the other hand it is extremely prominent in the macro limit, so that quantum evolution effectively appears like classical evolution on trajectories which obey Newtonian dynamics. The quantum to classical transition is naturally explained, and there is no longer any need for a measuring apparatus, to explain the results of measurements.

The most interesting thing about collapse models is not that they are necessarily correct, but rather that they are experimentally testable and that in principle they make predictions which are different from those of quantum mechanics. In the micro regime, the rate constant is so small that the models are indistinguishable from Schr\"{o}dinger evolution and hence make essentially the same experimental predictions as quantum mechanics. In the macro regime the predictions are the same as that of classical mechanics. It is in the 
in-between mesoscopic regime - the thirteen orders of magnitude alluded to at the beginning of the 
article -  that the model predictions markedly differ from that of quantum mechanics. The principle effect is that in this range  the lifetime of a quantum  superposition is neither too large nor too small, but in a range 
suitable for experimental detection. Thus if a mesoscopic object, having a mass of say a billion a.m.u., is prepared in a superposed state by passing it through a diffraction grating, then according to collapse models this superposition  will decay before the particle reaches the detecting screen, and hence no interference pattern will be seen. If this happens, it of course violates quantum mechanics, and is evidence for collapse models. Experiments of this nature form the subject of matter wave interferometry, and they have played a very important role in constraining collapse models and putting bounds on the rate constant $\lambda$ ~\cite{Hornberger2011review}. A great technological challenge is to eliminate`impurities' such as ambient radiation and gas which cause environmental decoherence, and mask and mimic the loss of superposition caused by collapse models. The largest objects for which superposition has been verified through interferometry have a mass of about $10^{5}$ a.m.u. and this puts an upper bound on $\lambda$ of about $10^{-5}$ sec$^{-1}$ ~\cite{RMP:2012}.

A different class of experiments which are becoming important in testing and constraining collapse models have to do with a side effect of these models. Namely, the stochastic process which introduces randomness in the dynamics also causes stochastic heating of the affected quantum particle, and hence a very tiny violation of energy momentum conservation ~\cite{Adler3:07}. The fact that such a violation has not been observed in laboratory experiments and in astronomical observations puts powerful bounds on $\lambda$, the strongest current bound being that $\lambda < 10^{-8}$ sec$^{-1}$ ~\cite{Vinante:2016}. 
Various new experiments have been proposed to test the effects of stochastic heating ~\cite{PEARLE5,Bera:15, Bahrami:2014a,Goldwater:2015}. 
Eventually, in order to verify or rule out collapse models, experiments must push this bound all the way down to $10^{-17}$, below which value collapse models may not be able to solve the measurement problem, and other explanations such as many worlds and Bohmian mechanics would start to appear more favorable. 

Collapse models do indeed have some limitations, which call for their better theoretical understanding. The models are purely phenomenological in nature, having been proposed with the express purpose of solving the quantum measurement problem. The mathematical structure of the stochastic nonlinearity is designed so as to give rise to the Born probability rule. In that sense the models do not predict or prove the Born rule; rather they have the Born rule built into them. Thus the question as to what is the fundamental origin of the probabilities still remains uanswered. [The same is true of the many worlds picture, and of Bohmian mechanics as well.] We really do not know what is the cause of this randomness. Why should there be in nature this stochastic noise field which these models employ? 

Two ideas which bear on this question in a serious way deserve mention. One is that this stochasticity has to do with gravity and spacetime structure. Gravitational fields are produced by macroscopic bodies, and the latter obey the uncertainty principle of quantum mechanics. It seems plausible [though not fool-proof] that this introduces an uncertainty in the produced gravitational field, and hence fluctuations in the spacetime geometry. This might be the source of randomness sought for by collapse models. It is then natural to ask how these fluctuations in the geometry affect the motion of a quantum particle which obeys Schr\"{o}dinger evolution? Various model studies have shown that spacetime fluctuations produce gravitationally induced decoherence of the wave function, with the effect becoming more prominent as the mass of the quantum particle is increased ~\cite{Karolyhazi:66,Karolyhazi:86, Karolyhazy:74, Karolyhazy:90, Karolyhazy:95, Karolyhazy:1982, Frenkel:77, Frenkel:90, Frenkel:95, Frenkel:2002, Frenkel:97, Diosi:87, Diosi:07, Diosi:89, Diosi:87a, Penrose:96, Penrose:98, Penrose:00, Diosi:84,Bernstein:98,giulini2011gravitationally,Harrison:2003,Moroz:98,Ruffini:69,Giulini2012,Giulini2013,Hu2014,Anastapoulos:2014, Bahrami:2014, Colin2014,Bera:2015,Bera:2015b,Bera:2016a,Derakhshani:2014}.
    While these results are very encouraging, they do not yet provide a collapse model. Gravity can cause decoherence, but it is not yet clear how (if at all) it causes collapse of the wave function (selection of one of the various outcomes) and how it explains the Born probability rule. The conceptual status of gravity in such models is also not very clear: is gravity classical, quantum, semiclassical, or something else? Nonetheless, since we know that gravity exists, it is very promising to investigate if it is the source of the nonlinear stochasticity in collapse models.

The second idea for a fundamental origin of collapse models is to consider if quantum theory is an approximation to a deeper underlying theory, and if the nonlinear stochastic modification arises as a higher order correction to the leading approximation. That quantum theory should perhaps be formulated differently, starting from some fundamental principles, is already indicated by the extreme dependence of the current formulation of the theory on its own classical limit. This is the essence of the theory of Trace Dynamics [TD], developed by Adler and collaborators ~\cite{Adler:94, Adler:04, Adler-Millard:1996, Adler:06a}.
  TD is the classical dynamics of matrices $q_r$ whose elements can either be odd grade [fermionic sector F] or even grade [bosonic sector B] elements of Grassmann numbers. The Lagrangian in this dynamics is defined as the trace of a polynomial function of the matrices and their time derivatives. Lagrangian and Hamiltonian dynamics can then be developed in the conventional manner. In TD, the matrix-valued configuration variables $q_r$ and their conjugate momenta $p_r$ all obey arbitrary commutation relations amongst each other.  However, as a consequence of a global unitary invariance of the dynamics there occurs in TD an important conserved charge, known as the Adler-Millard charge
\begin{equation}
\tilde{C} = \sum_B [q_r,p_r] -\sum_F \{q_r,p_r\} 
\end{equation}
whose existence is central to the subsequent development of the theory.

Assuming that one is not examining the dynamics exactly, one develops an equilibrium statistical thermodynamics for the classical dynamics described by TD. If one considers a sufficiently large system [of many, many 
particles, each particle being a matrix, as if there were a gas of matrices], the `system point' can in the long run be assumed to scan all of phase space. The phase space probability distribution achieves equilibrium [i.e. a uniform distribution over phase space]. The equilibrium distribution can be determined by maximising the entropy, as is done in statistical mechanics.  The equipartition of the Adler-Millard charge leads to certain Ward identities, which in turn lead to the important result that thermal averages of canonical variables obey quantum dynamics and quantum commutation relations. In particular, the emergent $q$ operators commute with each other, and so do the $p$ operators. This is how quantum theory is seen as an emergent phenomenon. The quantum state satisfying the Schr\"{o}dinger picture is recovered as usual, by implementing a transition from the Heisenberg picture to the Schr\"{o}dinger picture. 
TD is a classical deterministic theory, and time evolution of the matrices is described in the standard way, with respect to a flat Minkowski spacetime background. However, TD is {\it not} a hidden variable theory, because the matrix variables exist at a distinctly different underlying level, as compared to the quantum theoretical degrees of freedom, with the latter arising only upon statistical
coarse-graining, in the conventional sense of statistical mechanics. Hence the arguments of Bell's theorem against local hidden variable theories do not apply to TD.

Furthermore, if one considers the inevitable statistical fluctuations of the Adler-Millard charge about equilibrium, this leads to a collapse model type modification of the nonrelativistic Schr\"odinger equation. 
These fluctuations are the sought for source of randomness.
One does not understand TD well enough to uniquely predict the modified theory. In particular one still does not have a proof of the origin of Born probability rule in TD, but TD is perhaps the only theory to date, apart from gravity, which provides a fundamental explanation for randomness, by way of the statistical fluctuations. The collapse models, which are highly successful phenomenologically, are one possible modification admitted by TD. 
The modification, ignorable for microscopic objects but significant for large objects, solves the quantum measurement problem and leads to emergent classical behavior in macroscopic systems.  The fluctuations of the conserved charge about its equilibrium value carry crucial information about the arbitrary commutation relations amongst the configuration variables and their momenta in the underlying TD.

Coming back to collapse models, another of their limitations is that they are non-relativistic. Various attempts to construct relativistic collapse models face difficulties, a feature shared also by Bohmian mechanics. Perhaps this is an indicator that collapse may not be compatible with special relativity, especially in the light of quantum non-locality related issues which we discuss in a subsequent section below.

We take this occasion to mention that macroscopic quantum states such as superconductors and Bose-Einstein condensates, which are made from superposition of internal degrees of freedom, do not invalidate collapse models. The constraints on the rate constant  $\lambda$ from such systems are rather weak.

We conclude this section by noting that important theoretical and experimental progress is currently being made on the quantum measurement problem, and on removing this aspect of the dependence of the theory on its classical limit. We can expect some exciting developments in this problem in the coming decade or so.


\section{The problem of time in quantum theory}
\label{sec:3}
The time in quantum theory is part of a classical spacetime geometry, which geometry is produced by macroscopic bodies, which in turn are a limiting case of quantum objects, whose evolution is described with respect to this very time! It is evident that in order to avoid this self-reference there ought to exist an equivalent reformulation of quantum theory, which does not refer to classical time. This problem is no less severe than the measurement problem, but somehow it gets far less attention, if any at all.

In searching for such a reformulation we are guided by the assumption that such a reformulation should also throw light on the quantum measurement problem. After all both these problems arise from the dependence 
of quantum theory on its classical limit, and a common explanation is not implausible. We are also motivated by the fact that Trace Dynamics already seeks to obtain quantum theory, and its stochastic nonlinear modification, from underlying deeper principles, albeit while retaining the classical structure of spacetime. From our point of view however, as expressed above, the dependence of quantum theory on classical time seems to be a limitation, and we have made preliminary attempts to extend TD to remove the dependence on classical time. This is still work in progress and we summarize below what has been understood so far ~\cite{Singh:2012,Lochan-Singh:2011,Lochan:2012}.

To achieve a formulation of quantum theory without classical time, we first generalized Trace Dynamics so as to make space-time coordinates also into operators. Associated with every degree of freedom there now are coordinate operators  $(\hat{t}, \hat{\bf x})$ with arbitrary commutation relations amongst them. From these we construct a Lorentz invariant line-element $d\hat{s}^2$, and we define the important notion of Trace time $s$ as follows:
\begin{equation}
ds^2 = Tr d\hat{s}^2 \equiv Tr[d\hat{t}^2 - d\hat{x}^2 - d\hat{y}^2 - d\hat{z}^2]
\label{nsr}
\end{equation}
A Poincar\'e invariant dynamics is constructed, in analogy with ordinary special relativity, and in analogy with TD, but with the difference that evolution is now defined with respect to trace time $s$. The theory, as before,  admits a conserved Adler-Millard charge, and the degrees of freedom now involve bosonic and fermionic components of space-time operators as well. Because the space-time operators have arbitrary commutation relations, there is now no point structure or light-cone structure, nor a notion of causality, although the line-element is Lorentz invariant.

From this generalized TD, we constructed its equilibrium statistical thermodynamics, as before. The equipartition of the Adler-Millard charge results in the emergence of a generalized quantum dynamics [GQD] in which evolution is with respect to the trace time $s$, and the thermally averaged space-time operators $(\hat{t}, \hat{\bf x})$ are now a subset of the configuration variables of the system. It is significant that these averaged operators commute with each other. This is the originally sought after reformulation of quantum theory which does not refer to classical time. In the non-relativistic limit we recover the generalized 
Schr\"odinger equation
\begin{equation}
i\hbar \frac{d\Psi(s)}{ds} = H\Psi (s)
\label{gqd}
\end{equation}

To go beyond special relativity, one must invoke an operator structure for the spacetime metric. Here the program runs into difficulties. It has been argued by Adler that the metric must retain its classical [non-operator] structure in TD. If a way can be found around this, we expect the development to proceed along the following lines.

To demonstrate the equivalence of the reformulation [GQD] with standard quantum theory, one must first explain how the classical Universe, with its classical matter fields and ordinary space-time, emerges from the GQD in the macroscopic approximation. Like in TD, one would next allow for inclusion of stochastic fluctuations of the Adler-Millard charge, in the Ward identity. This should result in a non-linear stochastic Schr\"odinger equation, but now with important additional consequences. One considers the situation where matter starts to form macroscopic clumps (as for example in the very early universe, right after the Big Bang). These stochastic fluctuations become increasingly significant as the number of degrees of freedom in the clumping system  increases. As in collapse models, these fluctuations result in macroscopic objects  being localized, but now not only in space, but  in time as well! This means that the time operator associated with every object becomes classical (i.e. it takes the form: a $c$-number times a unit matrix).

The localization of macroscopic objects is thus accompanied by the emergence of a classical space-time. This is in accordance with the Einstein hole argument: classical matter fields and the metric they produce are required to give physical meaning to the point structure of spacetime.  If, and only if, the Universe is dominated by macroscopic objects, as is the case in today's Universe, can one also talk of the existence of a classical space-time. When this happens, the trace proper time $s$ can be identified with classical proper time. After the Universe reaches this classical state, it sustains this state, because of the continuous action of stochastic fluctuations on macroscopic objects, thereby simultaneously achieving  the existence of a classical space-time geometry. Since the underlying generalized TD is Lorentz invariant, the emergent classical  space-time is locally Lorentz invariant too. However there is a key difference: unlike in the underlying theory, now the light-cone structure, and causality, are emergent features, because the space-time coordinates have become $c$-numbers now.

Irrespective of this pre-existing classical spacetime background, a microscopic system in the laboratory is described at a fundamental level in terms of its own non-commutative space-time (\ref{nsr}), via the generalized TD associated with it. Subsequent to coarse-graining, this results in the system's GQD (\ref{gqd}) with its trace time. If we assume  that stochastic fluctuations can be ignored, this GQD has commuting $\hat t$ and $\hat {\bf x}$ operators. These, because of their commutativity, can be mapped to the $c$-number $t$ and ${\bf x}$ coordinates of the pre-existing classical universe, and trace time can then be mapped to ordinary proper time. This is hence a mapping to ordinary special relativity, and one recovers standard relativistic quantum mechanics in this way, as well as its non-relativistic limit. If this program can be fully implemented, it will establish  as to how standard quantum theory is recovered from the reformulation which does not depend on classical time.

Thus in our scenario the problem of time and the problem of measurement are related to each other. If one starts from a formulation of quantum theory which does not have classical time, then, in order to recover classical time and spacetime geometry from it, one must also recover from this formulation the macroscopic limit of matter fields. This is because classical geometry and classical matter fields go hand in hand. And to recover the classical limit for macroscopic objects is the same thing as solving the measurement problem. Because the latter problem can be restated as: why are macroscopic objects not found in superposition of position states?  The measurement problem is a subset of the larger question: how does the classical structure of spacetime and matter emerge from an underlying quantum theory of spacetime and matter?

\section{Quantum non-locality and space-time structure}
\label{sec:4}
The essence of the EPR paradox is that measurement on one part of a quantum system instantaneously influences another part of the same (correlated) quantum system, even if the two sub-systems are 
space-like separated.  To Einstein, this suggested that quantum theory is incomplete.
However, experimental measurements on entangled quantum states indeed demonstrate non-local correlations and indeed suggest the existence of an acausal action at a distance across space-like separated regions. This has been confirmed by increasingly precise loophole free tests of violation of Bell's inequalities by quantum systems. Although such correlations cannot be used for superluminal signaling, the acausal nature of the influence suggests the possibility of a conflict with special relativity and Lorentz covariance. This so-called spooky action at a distance has been debated extensively, but numerous investigations over decades have not provided a satisfactory resolution of the issue. On the other hand, a remarkable experiment shows that even if one  assumes that the influence travels causally in a hypothetical privileged frame of reference, its speed would have be  at least four orders of magnitude greater than the speed of light. That in itself could lead to problems with special relativity. Furthermore, we have seen above that attempts to construct relativistic versions of collapse models run into difficulties. To us this is possibly a signal that wave function collapse is not compatible with classical spacetime structure [light-cones, and causality].

A possible resolution might come from the underlying noncommutative structure of spacetime that we have proposed, and which was discussed above, in the context of the problem of time. It may well be that trying to describe collapse from the viewpoint of ordinary spacetime is not the right thing to do, when one goes over from the absolute Newton time of non-relativistic quantum mechanics, to the relative time of special relativity ~\cite{Bera:2016}. Collapse is perceived as instantaneous in terms of ordinary time, but there is nothing to say that this is the correct time to use. We have to pay heed that this classical time is external to quantum theory. 

Let us go back to the Generalized Quantum Dynamics [GQD] where evolution of the quantum system is described with respect to trace time $s$. Before the measurement takes place, the stochastic fluctuations of the Adler-Millard charge can be neglected for the quantum system [since it is microscopic], and as we observed above, its GQD can be mapped to standard quantum theory. However, when the measurement is done, the collapse inducing stochastic fluctuations in the space-time operators $\hat{t}, \hat{\bf x}$ associated with the quantum system become significant. These operators now carry information about the arbitrary commutation relations of the underlying generalized TD and they no longer commute with each other. This implies that they cannot be mapped to the ordinary space-time coordinates of special relativity.   Here, simultaneity can only be defined with respect to the trace time $s$, and there is no special relativistic theory of wave function collapse. In this picture, collapse and the so-called non-local quantum correlation   takes place only in the non-commutative space-time (\ref{nsr}), which lacks point structure, lacks light-cone structure, and is also devoid of the notion of distance. Therefore one can only say that collapse takes place at a particular trace time, which is Lorentz invariant, and it is not physically meaningful to talk of an influence that has travelled, nor should one call the correlation non-local. In this picture the wave function does not know distance - it just is. We once again see that getting rid of classical spacetime from quantum theory removes another one of its peculiarity, the so-called spooky action at a distance.

If, as is conventionally done, one tries to view and describe the measurement on the entangled quantum state from the view-point of the Minkowski space-time of special relativity, the process inevitably appears acausal and non-local. However, such a description should not be considered valid, because there is no map from the fluctuating and noncommuting 
${\hat t}, \hat{\bf x}$ operators to the commuting $t$ and ${\bf x}$ coordinates of ordinary special relativity. No such map exists in the non-relativistic case either. However, in the non-relativistic case,  because there is an absolute time, it becomes possible to  model the fluctuations as a stochastic field on a given space-time background, as is done in collapse models, and collapse is instantaneous in this absolute time; however it does not violate causality.

We see that while on the one hand the problem of time is related to the measurement problem, on the other hand, the resolution of the time problem can alleviate the mysterious nature of quantum non-locality. It will be interesting to investigate if one can make an experimental proposal to verify if noncommutative spacetime is indeed the way to understand the spooky action at a distance.

Undoubtedly, much more work needs to be done, to put the ideas of the present and the previous section on a firm footing.

\section{Implications for a quantum theory of gravity?}
\label{sec:5}
The three problems that we have discussed here could all be called a conflict between quantum theory and general relativity. The measurement problem has to do with the classical [as opposed to quantum] nature of macroscopic objects. These objects are intimately tied up with spacetime geometry through the laws of general relativity. To the extent that quantum theory does not explain the properties of macroscopic objects, it maybe said to be in conflict with general relativity. The problem of time is a direct conflict of course, because quantum objects do not produce a classical spacetime geometry. And also, quantum non-locality does not seem consistent with classical spacetime structure.

Given all this, should we aim to construct a quantum theory of gravity by `quantizing' classical general relativity? It seems rather unnatural to do so. It is a fine thing to quantize other fundamental forces, because they take spacetime structure as given, and because they do not face the kind of conflict that gravity faces with quantum theory. By quantizing general relativity, we seem to violate the rules of the game. There is this classical spacetime sructure whose existence is pre-assumed while writing down the quantum rules: how can these rules then be applied to that very structure? It does not seem a logical thing to do, and there is no guarantee that the correct quantum theory of gravity will emerge in this way.

Rather, we see pressing reasons - measurement problem, time problem, non-locality - which suggest the need to modify both quantum theory and spacetime structure, when one starts trying to resolve the conflict between classical and quantum. We should not quantize gravity; rather there is an underlying theory - perhaps a combination of noncommutative geometry and Trace Dynamics as suggested here - or something else, from which both quantum theory and gravitation are emergent. Gravitation emerges in the full classical limit, when both matter and gravity are treated classically. Quantum theory emerges, upon coarse graining the underlying theory, when only the gravity sector is treated classically. It maybe that this underlying theory is arrived at by demanding that physical laws be covariant under general coordinate transformations of 
non-commuting coordinates, thus bringing together the element of general covariance from relativity, and the element of noncommutativity from quantum theory. Given the nonlinearity of gravitation, it seems rather unlikely that the principle of quantum linear superposition can survive such a union! 

\bigskip
\begin{acknowledgement}
It is my pleasure to thank my collaborators  on these topics - Angelo Bassi, Hendrik Ulbricht, Saikat Ghosh, Shreya Banerjee, Srimanta Banerjee, Sayantani Bera, Suratna Das, Sandro Donadi, Suman Ghosh, Kinjalk Lochan, Seema Satin, Priyanka Giri, Navya Gupta, Bhawna Motwani, Ravi Mohan and Anushrut Sharma.
\end{acknowledgement} 

\bigskip

\bigskip


\bibliographystyle{spphys}
\bibliography{biblioqmtstorsion}

\end{document}